\documentclass[manuscript]{acmart}

\AtBeginDocument{%
  \providecommand\BibTeX{{%
    \normalfont B\kern-0.5em{\scshape i\kern-0.25em b}\kern-0.8em\TeX}}}

\copyrightyear{2022}
\acmYear{2022}
\setcopyright{rightsretained}
\acmConference[EAAMO '22]{Equity and Access in Algorithms, Mechanisms, and Optimization}{October 6--9, 2022}{Arlington, VA, USA}
\acmBooktitle{Equity and Access in Algorithms, Mechanisms, and Optimization (EAAMO '22), October 6--9, 2022, Arlington, VA, USA}
\acmDOI{10.1145/3551624.3555290}
\acmISBN{978-1-4503-9477-2/22/10}

\usepackage{xcolor}
\usepackage{wrapfig}

\begin{document}

\title{Power to the People? Opportunities and Challenges for Participatory AI}

\author{Abeba Birhane}
 \authornote{This work was conducted during Abeba Birhane's internship at DeepMind.}
\affiliation{%
  \institution{Mozilla Foundation \& University College Dublin}
  \country{Ireland}
}
\email{abeba.birhane@ucdconnect.ie}
\orcid{0000-0001-6319-7937}

\author{William Isaac}
\affiliation{%
  \institution{DeepMind}
  \country{UK}
}
\email{williamis@deepmind.com}
\orcid{0000-0002-1297-5409}

\author{Vinodkumar Prabhakaran}
\affiliation{%
  \institution{Google}
  \country{USA}
}
\email{vinodkpg@google.com}
\orcid{0000-0003-3329-2305}

\author{Mark Díaz}
\affiliation{%
 \institution{Google}
 \country{USA}}
\email{markdiaz@google.com}
\orcid{0000-0003-0167-9839}

\author{Madeleine Clare Elish}
\affiliation{%
  \institution{Google}
  \country{USA}}
  \email{mcelish@google.com}
  \orcid{0000-0002-9647-1178}

\author{Iason Gabriel}
\affiliation{%
  \institution{DeepMind}
  \country{UK}}
\email{iason@deepmind.com}
\orcid{0000-0002-7552-4576}
 
\author{Shakir Mohamed}
\affiliation{%
  \institution{DeepMind}
  \country{UK}}
\email{shakir@deepmind.com}
\orcid{0000-0002-1184-5776}

\renewcommand{\shortauthors}{Birhane, et al.}

 \begin{abstract}
Participatory approaches to artificial intelligence (AI) and machine learning (ML) are gaining momentum: the increased attention comes partly with the view that participation opens the gateway to an inclusive, equitable, robust, responsible and trustworthy AI. Among other benefits, participatory approaches are essential to understanding and adequately representing the needs, desires and perspectives of historically marginalized communities. However, there currently exists lack of clarity on what meaningful participation \textit{entails} and what it is \textit{expected} to do. In this paper we first review participatory approaches as situated in historical contexts as well as participatory methods and practices within the AI and ML pipeline. We then introduce three case studies in participatory AI. 
Participation holds the potential for beneficial, emancipatory and empowering technology design, development and deployment while also being at risk for concerns such as cooptation and conflation with other activities. We lay out these limitations and concerns and argue that as participatory AI/ML becomes in vogue, a contextual and nuanced understanding of the term as well as consideration of who the primary beneficiaries of participatory activities ought to be constitute crucial factors to realizing the benefits and opportunities that participation brings.  

\end{abstract}



\maketitle

\section{Introduction}

Artificial Intelligence (AI) has taken a `participatory turn', with the reasoning that \textit{participation} provides a means to incorporate wider publics into the development and deployment of AI systems. The greater attention to participatory methods, participatory design, and the emerging imaginary of participatory AI, follows as a response to the changing attitudes towards AI's role in our societies, in light of the documented harms that have emerged in the areas of security, justice, employment, and healthcare, among others~\citep{costanza2020design,muller2012participatory,van2015participatory,harrington2020forgotten,arana2021citizen,waycott2016not,weber2015participatory,majale2008employment}. The field of artificial intelligence is faced with the need to evolve its development practices---characterized currently as technically-focused, representationally imbalanced, and non-participatory---if it is to meet the optimistic vision of AI intended to deeply support human agency and enhance prosperity. Participation has a vital role to play in aligning AI towards prosperity, especially of the most marginalized, but requires a deeper interrogation of its scope and limitations, uses and misuses, and the place of participation within the broader AI development ecosystem. 

A growing body of work has shown the different roles and formats that participation can take in the development of AI, including: new approaches to technical development in NLP in healthcare~\citep{nekoto2020participatory, donia2021co}, in the development of alternative design toolkits and processes~\citep{katellToward2020, martin2020participatory}, and methods that range from structured interviews to citizens juries~\citep{balaram2018artificial}. In these cases, participation is meant to move beyond individual opinion to center the values of inclusion, plurality, collective safety, and ownership, subsequently shifting the relationship from one of designer-and-user to one of co-designers and co-creators. Participation is expected to lead to systems of self-determination and community empowerment. Yet, caution has been raised about the possibility of `participation-washing'~\citep{sloane2020participation}, where efforts are mischaracterized under the banner of participation, are weakly-executed, or co-opt the voice of participants to achieve predetermined aims. 

In this paper, we advance a view that participation should continue to grow and be refined as a key component of the AI development and deployment lifecycle as a means to empower communities, especially those at the margins of society that AI often disproportionately and negatively impacts. To achieve this through Participatory AI, greater clarity is needed on \textit{what} participation is, \textit{who} it is supposed to serve, \textit{how} it can be used in the specific context of AI, and \textit{how} it is related to the mechanisms and approaches already available. Our paper makes three contributions towards developing this deeper understanding of participatory AI.
Firstly, we situate the participatory process within its broader genealogical development. We develop this historical hindsight in section \ref{sect:history}, considering histories of participatory 
    development 
    as well as its colonial inheritances and its newer digital 
    forms. 
Secondly, we present three case studies in section \ref{sect:casestudies} taken from selected participatory projects to further concretize various forms of existing participatory work. 
we reframe participation by introducing a characterisation that allows the multiple forms of participation seen in practice to be compared in Appendix B.  
We then describe potential limitations and concerns of participation in section \ref{sect:limitations} and conclude in Section~\ref{sect:conclusion}.

\section{Genealogy of Participation}
\label{sect:history}
\begin{quote}
    “\textit{‘je participe, tu participes, il participe, nous participons, vous participez … ils profitent’} (in English: \textit{‘I participate, you participate, he participates, we participate, you participate … they profit’})”. \footnote{From a French poster by Atelier Populaire,  May 1968. V\&A Museum collections accession number E.784-2003 }
\end{quote}
This quote appears widely in works related to participation~\citep{arnstein1969ladder, chilvers2015remaking, kelty2020participant}---it poetically captures the cycles of enthusiasm and use of participation and participatory methods as a remedy for many problems of social life, but ending with a sense of disenchantment, exploitation, and with asymmetrical power dynamics in society left unchanged. We see the same enthusiasm for participation in AI at present, which renews this quote's relevance for the analysis of participatory approaches in AI. The quote points to the many historical lessons upon which new participatory undertakings can draw from, which is often absent in AI research; it also serves as a warning of one type of participatory outcome to avoid. 
To draw on this experience, in this section, we begin with a summary of participation's historical roots, look specifically at participation for technological development, and then review the current landscape of participatory AI.

\subsection{Historical Participation}
Over the last century, participatory approaches have risen globally to the fore across all sectors, including in international development, healthcare provision, decision-making about science, democracy, the environment, and social movements for self-determination, among others~\cite{chilvers2015remaking,groves2017remaking,dearden2008participatory,chambers1994origins, kelty2020participant, freire1996pedagogy}. This rise is driven by the multitude of benefits associated with participation. Participatory approaches, by engaging citizens in scientific, democratic or environmental decision-making, for example, enables these processes to become transparent, accountable, and responsive to the needs of those who participate. Participatory methods also establish a distinct opportunity to learn from and include knowledge created by the people directly affected by existing or new undertakings. When such collective knowledge is missing, its absence leads to failure leaving projects to be based solely upon technocratic or elite perspectives~\cite{ferguson1994anti,mitchell2002rule, toyama2015geek}. Moreover, at its best, participation leads to individual and collective empowerment as well as social and structural change via the cultivation of new skills, social capital, networks, and self-determination among those who contribute. This has the potential to make a sustained positive impact to the welfare and benefit of communities over time~\cite{sen2009idea,farmer2001community}.

The desire to unlock these benefits through novel forms of organization played a central role in the development of participatory approaches to research and decision-making, a trend that is most often traced back to the work of Scandinavian researchers in the 1970s and 80s~\cite{asaro2000transforming, chilvers2015remaking}. The ‘Scandinavian approach’ to participation is concerned primarily with the creation of ‘workplace democracy’ understood as a system of structured consultation and dialogue between workers and employees with the aim of giving workers greater control over wages and the allocation of tasks. Building upon this idea, participatory approaches have been used to countenance different kinds of response to the challenges posed to workers by technological innovation. 
As examples, the \textit{Scandanavian Collective Resource Approach} helps workers actively manage processes of technological adoption by promoting knowledge-sharing about new technologies, improving the ability of unions to negotiate collectively with employers, and identify mutually beneficial trajectories~\cite{asaro2000transforming,kraft1994collective}. The \textit{British Socio-Technical Systems Approach} to participation was developed to promote the notion of a \textit{systems science} where new technologies, the workers (with emphasis on their psychological and physical well-being), and the environment that they are embedded in, are held to be an interactive part of a larger system that needs to be collectively managed. The latter school of thought set out to promote workers' autonomy through their active participation in the design of socio-technical systems as a whole~\cite{ehn1987collective,asaro2000transforming}.

The power of participation has led to a proliferation of approaches, including the principle of maximum feasible participation~\citep{moynihan1969maximum}, and one of the most regarded uses in the process of participatory budgeting~\citep{cabannes2004participatory}. Today there are numerous tools and processes available for anyone to build on the established practices across a range of participatory methods, whether they include Delphi interviews, citizen's juries, role-playing exercises and scenarios,  workshops, force field analyses, or visual imagery (like those in the participatory development toolkit~\citep{kelty2017toolkit}), alongside the professionalization of participation through organisations like the International Association for Public Participation. 

Despite these advancements, historical analysis of the roots of participation reveal some of its failings and shortcomings. Long before calls for participation in the workplace, the notion of participation played a central role in the administration of the British empire. The ``Dual Mandate in British Tropical Africa'' colonial system of the 1920s was rooted in the co-optation of traditional rules and authority structures~\cite{lugard1922dual}. By establishing a hierarchical division of power that was enforced using the authority of local rules and chiefs, colonial projects claimed their legitimacy under the veneer of participation. They mandated local people to abide by colonial rules, turning participation in governance into a form of colonial power.

The risk that participation could simply mask uneven power relations, making it easier to perpetuate a dynamic that is fundamentally extractive remains a major concern. One of the most astute critique in this vein was raised by~\citet{arnstein1969ladder} that established the now widespread image of the ladder of participation that provided a linear model with which to characterise participation, from extractive to empowering. Although the ladder was powerful, the key critique~\citet{arnstein1969ladder} raised was one of power. And this critique has been extended further, labelling the fervour for participation in all sectors as a form tyranny, pushing out other approaches for equity that would be more appropriate while using participatory methods to facilitate an illegitimate and unjust exercise of power. As~\citet{arnstein1969ladder} writes:  ``... participation without redistribution of power is an empty and frustrating process for the powerless. It allows the powerholders to claim that all sides were considered, but makes it possible for only some of those sides to benefit. It maintains the status quo.''

\subsection{Participation for Technological Innovation}
Of relevance to machine learning research, are the specific roles that calls for participation have played in the context of computing and technological innovation. In the U.S., participatory design was widely adopted by large corporations creating office technologies in the 1970s and 80s~\cite{asaro2000transforming}. The key idea was to increase the quality of products, and to reduce time and costs, by bridging the gap between those doing the design (removed from day to day use) and those that were designed for by involving end-users in the process. Participation in this sense was primarily conceived of as technical work done by participants for the sake of economic efficiency; an aspiration that fit well with larger organizational goals. While this approach yielded some benefits for consumers via an improved product, the value of participation was limited: it did not need to benefit those engaged in co-design and was very unlikely to empower them.

One salient question centers upon whether participation in the technology development pipeline necessarily requires those involved to be actively engaged in the process. For those who focus on participation in the form of activism, movement-building and community initiatives, active  engagement is essential~\cite{chilvers2015remaking}. Yet, others  define participation more widely~\cite{barney2016participatory}, so that it encompasses types of incidental participation that arises by simply being part of an environment or process that involves others. This phenomenon is increasingly pronounced in digital media environments, e.g., having one's data harvested and used in AI development by virtue of ``participating'' in social media, sometimes referred to as mediated participation~\citep{kelty2020participant}, participation through the medium of technology. Yet this type of ``passive'' participation has increasingly been linked to privacy violations and surveillance~\cite{veliz2020privacy,zuboff2019age}.  Furthermore, to support the ideal of a ``participatory condition'' of society and technological development requires a degree of agency and intentionality~\cite{barney2016participatory, veliz2020privacy, kelty2020participant}. To participate, requires knowing that one is participating.
 
\subsection{The Emergence of Participatory AI}
\label{sect:current_ml_prticipation}

While the participatory mechanisms have served as a constant backdrop for the development of modern technologies, it's emergence within the context of artificial intelligence (AI) and machine learning based applications specifically have been relatively recent. Given its origins as a more speculative academic and industrial technological endeavor, the initial cycles of AI research largely missed the prior waves of participatory research that other technologies of comparable vintage (e.g. personal computing, internet networking, computer software). However, the shift away from logic-based AI systems towards more data-driven paradigms such Deep Learning~\cite{lecun2015deep} as well as new infrastructure for capturing and leveraging human-generated data (e.g. Amazon Mturk) prompted greater demand for ``non-expert'' participation in the construction of AI systems. 

One significant adoption of non-expert participation was in the construction of large scale benchmark datasets such as ImageNet~\cite{deng2009imagenet}, where the research team utilized over 49,000 workers from Amazon's mechanical turk (Mturk) platform across 167 countries~\cite{denton2021genealogy} to perform image recognition tasks to filter and validate the roughly 14 million images in the dataset. Despite this effort being quite broad in its ``participation'', the highly variable ethical and documentation standards~\cite{geiger2020garbage,denton2020bringing} for data enrichment or moderation tasks means that these contributors often fail to be discussed when publishing the final artefacts or protected by conventional research participant standards (e.g. Beneficence, Respect for Persons, Justice). Other area has been in the form of content moderation, where non-expert participants are used to review misinformation or graphic media to prohibit the display of harmful content on internet platforms (e.g YouTube, Facebook, Twitter) but also serve as labelled training data for machine learning classifiers deployed to expedite policy enforcement.  The proliferation of machine learning across multiple industries has further ingrained and expanded the general data enrichment and moderation paradigm, but the abuses and concentration of these forms of ``Ghost Work" in low income countries and peoples have also been extensively documented in recent years~\cite{irani2013turkopticon,gray2019ghost,roberts2019behind,mohamed2020decolonial}.

In parallel to the expansion of data enrichment and moderation, problematic applications of machine learning tools in high stakes domains such as criminal justice~\cite{lum2016predict, richardson2019dirty, bennett2018algorithmic}, healthcare~\cite{benjamin2019assessing,banerjee2021reading}, and hiring~\cite{raghavan2020mitigating, ajunwa2019auditing}
have prompted both researchers, civil society, and regulators to increasingly urge greater use participatory methods to mitigate sociotechnical risks~\cite{selbst2019fairness} not addressed by algorithmic adjustments or transformations. The recent upswing in participatory tools have varied in approach and point in the machine learning product lifecycle, including: auditing methods for machine learning based interventions~\cite{jo2020lessons,sambasivan2021everyone}, public consultation methods such as citizen juries~\cite{van2021trading, balaram2018artificial} or joint problem formulation~\cite{martin2020participatory}, information disclosures such as model cards~\cite{mitchell2019model,hutchinson2021towards} or datasheets~\cite{holland2018dataset,gebru2021datasheets}, and artefact co-development~\cite{lee2019webuildai,halfaker2020ores, sendak2020real}. A central tension of this this recent wave of ``participatory'' is whether these mechanisms should merely serve to aid in the refinement of relevant machine learning sytem or rather emphasize lived experience as a critical form of knowledge and employ experiential learning as a force for community empowerment and advance algorithmic equity ~\cite{katellToward2020,frey2019place} or ensure wider humanitarian or societal benefits~\cite{bondi2021envisioning,berditchevskaiaparticipatory}. The heavy influence of industry stakeholders calling for greater participation without resolving these tensions has led to concerns of `participation-washing' and calls for a greater need to focus on broader social structures and uneven power asymmetries~\citep{sloane2020participation, chan2021limits}, as well as the limits of participation in specific applications, such as healthcare~\citep{donia2021co}.

While the advancement of an emergent subfield of ``Participatory AI'' has its own critical questions and tensions which are important to further contexualize, the field needs to continue to reflect both its instrumental and broader purposes. The sections below focus exploring three specific areas:

\begin{enumerate}
  \item Standards: Despite many activities applying the label of participatory, there are yet no clear consensus on what minimum set of standards or dimensions one should use to assess or evaluate a given potential participatory mechanism. Though not an exhaustive list, 
 attributes such as the degree of \textbf{Reciprocity}, \textbf{Reflexivity}, and \textbf{Empowerment}, as well as the \textbf{Duration} of a task are applicable and salient considerations for 
 all participatory mechanisms. Please see the list of questions in Appendix A that further aid the reflexive process for those embarking on participatory activities. 
  \item Goals: There is no single unified vision of what Participatory AI tools are intend to achieve, but rather a bundle of overlapping goals and motivations. These include \textit{algorithmic performance improvements, process improvements, and collective exploration}. This is further explored in the Appendix B. While each of these objectives are participatory to some degree, the composition of the stakeholders and relative degree of influence 
  in ultimately shaping the development and impact of a given machine learning system vary significantly. Thus, researchers and developers must ensure that the forms of participatory mechanisms utilized align with the downstream technical, ethical and sociotechnical risks.
  \item Limitations: Invoking both lessons from history and contemporary cases, we will discuss some emerging limitations of utilizing participatory methods as a means of mitigating technical, ethical and sociotechnical risks. These include concerns of participatory mechanisms serving as a viable substitute for legitimate forms of democratic governance and regulation, co-optation of mechanisms in asymmetrical stakeholder settings, and conflation with other social goals such as inclusion or equity. See Section~\ref{sect:limitations} for more. 
\end{enumerate}

Below, we present three ``sites'' or case studies of Participatory AI across the machine learning lifecycle to explore the substantive areas outlined above. The decision to utilize a case study-based approach is intentional, aiming to provide a deeper understanding of the substantive questions in the context of existing or recent cases. Our hope is that this approach will lend a greater appreciation for the nuance and complexity implementing participatory mechanisms in this setting often presents to all the relevant stakeholders. Each case begins with a background description before presenting a contextual analysis. 
Through an investigation of existing participatory practices, we aim to offer a wider lens on the use of participatory mechanisms in the AI/ML context, where those goals can be attained through participatory approaches, and a clear understanding of potential limitations such that future efforts can hopefully fully realize the impact of meaningfully incorporating participation in AI/ML development and use.

\section{Three Case Studies in Participatory AI}
\label{sect:casestudies}
Participatory activities, process, and projects exist in a variety of forms. 
Within the \textit{AI for social good} field, for example, participatory activities are evoked as a means to improve AI systems that impact communities where, ideally, impacted groups take part as stakeholders through participatory design and implementation~\cite{bondi2021envisioning}. Participation has  also been instrumental in 
designing and building algorithms that serve communities for purposes such as operating on-demand food donation transportation service~\cite{lee2019webuildai} as well as for building tools, for instance, (Turkopticon) that allow low wage workers --- Amazon Mechanical Turkers, in this case --- to evaluate
their relationships with employers and support one another \cite{irani2013turkopticon}. Similarly, algorithmic tools that optimize worker well-being through participatory design and algorithmic tool building has been put forward by~\citet{leeparticipatory}. In collecting, documenting, and archiving, sociocultural data needed to train and validate 
machine learning systems, participatory approaches can be critical for representing the ``non-elites, the grassroots, the marginalized'' in datasets, as outlined by~\citet{jo2020lessons}. For justice oriented design of technology, participatory approaches provide the means for marginalized communities to organize grassroots movements, challenge structural inequalities and uneven power dynamics allowing communities to build the kind of technology that benefits such communities~\cite{costanza2020design}. Abstaining from participation can also be a form of participation in another related practice
, as shown in~\citet{waycott2016not}, who analysed older adults who refused to participate in technological intervention evaluation. Participatory projects, process, and objectives, therefore, are diverse and multifaceted. Below, we present three case studies that illustrate 
three various models of participation. 

\subsection{Case 1: Machine translation for African languages}

\paragraph{Description} Over 400 participants from more than 20 countries have been self-organizing through an online community. Some of the projects that have emerged from this community focus on the 
inclusion of traditionally omitted ``low-resourced'' African languages within 
the broader machine translation (MT) research agenda~\citep{nekoto2020participatory}. The projects sought to ensure that MT work for low-resourced 
languages is driven by communities who speak those languages. There were no prerequisites placed on participation or fixed roles assigned to participants but rather the roles emerged organically and participatory input was sought at every level of the process from language speaking to model building and participants moved fluidly between different roles. The research process was defined collaboratively and iteratively. Meeting agendas were public and democratically voted on. Language data was crowd-sourced, annotated, evaluated, analyzed and interpreted by participants (from participants). The specific project on MT for African languages 
 produced 46 benchmarks for machine translation from English to 39 African languages and from 3 different languages into English~\citep{nekoto2020participatory}. 

\paragraph{Analysis} 

The MT project by the Masakhane NLP community illustrates a grassroots (or bottom up) attempt at using participatory mechanism to build new systems improve the underlying performance of existing NLP systems through the inclusion of traditionally under-resourced African languages.   
The Masakhane MT project sought to 
increase the degree of empowerment for the stakeholders involved in the project. In this context, \textbf{Empowerment} reflects the degree of impact the participants have in shaping the participatory relationship, and ultimately the project or product. An empowering process is one that is often~\textbf{Reciprocal}: it is bi-directional, emergent, and dynamic, and one where core decisions or artefacts are informed by active participation rather than one based on the idea of placating a community or notions of saviourism.  For example, in this case, the 
idea is to not only crowd-source activities such as crowd-sourcing of language data, participant evaluation of model output, and production of benchmarks but also to create and foster 
avenues and forums to veto or voice objections.

Having said that, although the project is built on the idea that MT for low-resourced languages should be done by the language speakers themselves, for language speakers, based on community values and interests, it is still possible to see how the research, datasets and tools may be co-opted and monopolized by commercial actors to improve products or models without supporting the broader grassroots effort or the community's interests or needs. As a result the primary beneficiaries of participatory data sourcing may not be speakers of `low-resourced' languages but actors with access to such sufficient data and compute resources, thus gaining financial benefits, control and legitimacy off of such participatory efforts. 

\subsection{Case 2: Fighting for Māori data rights} 

\paragraph{Description}
Through participatory initiatives that took place over 10 days in 2018 as part of the \textit{Te Hiku} NLP project, the Māori community in New Zealand both recorded and annotated 300 hours of audio data of the \textit{Te Reo Māori} language~\citep{maori21}. This is enough data to build tools such as spell-checkers, grammar assistants, speech recognition, and speech-to-text technology. However, although the data originated from the Māori speakers across New Zealand 
and was annotated and cleaned by the Māori community itself, Western based data sharing/open data initiatives meant that the Māori community had to explicitly prevent corporate entities from getting hold of the dataset. The community thus established the Māori Data Sovereignty Protocols~\cite{rainie2019indigenous} in order to take control of their data and technology. Sharing their data, the Māori argued, is to invite commercial actors to shape the future of their language through tools developed by those without connection to the language. By not sharing their data, the Māori argue they are able to maintaining their autonomy and right to self-determination. They insist that, if any technology is to be built using such community sourced data, it must directly and primarily benefit the Māori people. Accordingly, such technology needs to be built by the Māori community itself since they hold the expert knowledge and experience of the language.   

\paragraph{Analysis}

The Māori case study is an illuminating example that brings together 
participatory mechanisms as means for methodological innovation while 
offering reciprocity to the relevant stakeholders. 
It is a process that prioritizes the \textit{net} benefit of participants, especially those disproportionately impacted by oppressive social structures, who often carry the burdens of negative impacts of technology \cite{mohamed2020decolonial,mhlambi2020rationality,couldry2021decolonial,birhane2021algorithmic} and reflecting a fair or proportionate return for the value of the participatory engagement. This is of particularly importance when seeking to utilize participatory mechanisms to achieve methodological innovation, or where the process yields unique insights that can inform new or innovative technological artefacts (as opposed to a means to achieve a particular pre-determined technical objective).

Because the data originates from the language speakers themselves and is annotated and cleaned by the Māori community, existing laws around data sovereignty~\cite{rainie2019indigenous} often require that those communities are key decision makers for any downstream applications. In this case, the Māori are committed to the view that any project created using Māori data must directly benefit and needs to be carried out by the Māori themselves. This high degree of reciprocity between stakeholders lead to a case where the needs, goals, benefits and interests of the community is central to participatory mechanism itself. This case study goes further than others 
by providing avenues for foregrounding reciprocity and refusal 
(when they are not aligned with 
the participants values, interests and benefits).

\subsection{Case 3: Participatory dataset documentation}

\paragraph{Description}
A team of researchers put forward participatory methods and processes for dataset documentation --- \textit{The Data Card Playbook} --- which they view as the route to creating responsible AI systems~\footnote{https://pair-code.github.io/datacardsplaybook/}. According to the team, the Playbook can play a central role in improving dataset quality, validity and reproducibility, all critical aspects of better performing, more accurate, and transparent AI systems. The Playbook comprises of three components -- Essentials, Module one, and Module two -- all activities supplemented by ready to download templates and instructions. These participatory activities cover guidance ranging from tracking progress, identifying stakeholders, characterizing target audiences for the Data Card, to evaluate and fine-tune documentation, all presented and organized in a detailed and systematic way. The Playbook aims to make datasets accessible to a wider set of stakeholders. The Playbook is presented as a people-centered approach to dataset documentation, subsequently, with the aim of informing and making AI products and research more transparent.  

\paragraph{Analysis}

This case study encapsulates the kind of participatory activities that support participation as a form of algorithmic performance and/or dataset quality improvement. 
An indirect benefit of this approach is that mechanisms designed to explore the space of a given artefact or process inevitably offer a potential for \textbf{Reflexivity}, critical evaluation and meaningful feedback. 
Reflexivity as part of a participatory process is a critical element for improving trust between stakeholders and conveying a sense legitimacy of the ultimate artefact.

However, because the central drive for these specific participatory practices are motivated by objectives such as dataset quality improvement, the participants are assigned pre-defined roles and very clear tasks. Dataset quality and transparent dataset documentation indeed impact the performance and accuracy of AI systems, which can all play a role in the larger goal of fair, inclusive, and just AI systems. Nonetheless, this form of participation focuses on fine-grained activities that come with pre-defined goals and objectives means that there is little room (if at all any) for co-exploring, co-creating, and/or negotiating the larger objectives, reflections, and implication of AI systems. There is no guarantee that an AI system that is created using improved and better datasets with the help of the Data Card Playbook cannot be used in applications that harm, disenfranchise, and unjustly target the most marginalised in society. Computer vision systems that are used to target refugees or marginalized communities in any society, for example, result in a net harm to the targeted regardless of their improved performance. Participation for algorithmic performance improvement is not necessarily equipped to deal with such concerns.

\section{Limitations and Concerns}
\label{sect:limitations}
Like any method, participation has limitations, and we briefly explore these here, and also refer to the large body of work in these topics~\citep{chilvers2015remaking, kelty2020participant, cooke2001participation, mackenzie2012value, barney2016participatory, mansuri2012localizing}. Effective participation should serve specific purposes and should not be conflated with other tasks and activities, such as consultation, inclusion, or labour. Moreover, participation cannot be expected to provide a solution for all concerns, and is not a solution for all problems. When used in considered ways, participation can be an important tool in the responsible development of AI. We consider here the role of participation in relation to democracy, its conflation with other activities, concerns on cooptation of participatory activities, the challenges of measuring the effectiveness of participatory efforts, and the challenges of balancing expertise and incentives. 

\textbf{Democratic governance.} In democratic societies, it is useful to think of democracy as an apparatus that responds to the right of citizens to determine the shape of practices that govern their lives. Consequently, participation is not the best mechanism for decisions/values/norms that are better decided and codified by democratic institutions, governance and laws. In a democratic system, participants are rights-holders: the people to whom decision-makers are accountable, and the body in which authority is ultimately vested. This distinction is important when an undertaking involves matters of significant public concern or modalities of operation, such as the coercive enforcement of law, that require stronger forms of validation and legitimacy~\citep{gabriel2022toward}. Participatory activities convened by private actors or parallel institutions, cannot stand in for democratic politics, and participatory AI should not aspire to do so or be perceived to meet this function.

\textbf{Conflation with other activities.} By acknowledging participation's limitations, we can refine what it does and does not entail. As one example, inclusion is often conflated with participation~\citep{quick2011distinguishing}. Being included might have practical consequences on the ability of people and groups to engage in participatory processes. But inclusion is not necessarily participation, since any individual can be included in a group, yet not participate, e.g., by never voting, writing, or acting. Inclusion is then related, but in some ways different from participation, and needs its own attention, which also depends on an understanding of any systemic and social barriers in place (e.g., racism, patriarchy, wealth exclusion). Attempts to include can also be exclusionary. When we invite people to participate it is never everyone: some people are always excluded. Typically those excluded are the very worst-off, those with low literacy, 
those who do not have the time to seek out participatory opportunities, are not members of the right networks, etc~\citep{spivak2003can}. At other times exclusion is needed for safe, open participatory action. And the purposeful abstention, collective refusal, dissent, or silent protest are themselves forms of participation (e.g., as illustrated by the 
Maori data rights case study). 

\textbf{Cooptation.} Concerns remain of participation becoming a mechanism of cooptation. Specific concerns are raised through current economic and capitalist models that seek to dissolve social relations and behaviours into commodities that are then open to monetization~\citep{zuboff2019age, couldry2019costs}. 
The history of colonial tactics showed how traditional participatory structures were co-opted to claim legitimacy. The case study on machine translation for African languages raises related concerns, where the efforts of grassroots participatory actions, and their data-sharing initiatives, leaves opens the door for cooptation, where corporate actors can use the results of participatory efforts towards corporate benefits. The potential for corporate actors to capitalize on such efforts and build products that maximize profits, 
with little benefit to communities remains open. In such circumstances, not only are those that participated disempowered, but corporations then emerge as the legitimate arbiters of African languages and subsequently language technology.

\textbf{Effectiveness and Measurement.} One core concern with participatory methods is that it is difficult to measure and provide attribution to the positive benefits of participation. The type of participation described in Appendix B are likely to result in benefits (though real) that are gradual and intangible, e.g., a sense of empowerment, knowledge transfer, creation of social capital, and social reform. In particular, participation that enables in-depth understanding and iterative co-learning can defy quantification and measurement. Investing in such types of participation may appear wasteful when outcomes are measured using blunt instruments such as cost-benefit analysis, and it could instead be the limitations of the metrics we use to evaluate participatory approaches that are an obstacle to the effective use of participatory approaches. 
The problem of measurement of impact and a general monitoring, evaluation and learning (MEL) framework is generally difficult, so also points to areas for further research to effectively make participatory methods part of regular practice~\cite{gabriel2017effective}.
 
\textbf{Expertise and Incentives.} One aim of participatory methods is to spread knowledge about technical systems and their impacts. This involves the knowledge of technical experts, but also the local knowledge embedded in the lived-experience of communities. There is an epistemic burden on all stakeholders in the participatory process to gather enough information to reason, questions, act or decide~\citep{pierre2021getting, scheall2021priority}. The need then to always learn and gather information requires participatory approaches that occur at different time frames, over various duration, and with different groups. Participation necessitates an assessment of the incentives involved, which can become distorted by epistemic burden, fundamentally affecting the participatory process. Put simply, participatory methods cannot rely on simplified assumptions about the reasons people have for engaging in a participatory process. This returns to the need to challenge uneven power distributions and oppressive social structures, as well as the ways that `community' itself can hide power dynamics.

\section{Conclusion}
\label{sect:conclusion}
To characterise AI as participatory is to acknowledge that the communities and publics beyond technical designers have  knowledge, expertise and interests that are essential to the development of AI that aims to strengthen justice and prosperity. Participation is then an essential component in achieving these aims, yet hype and misunderstanding of the participation's role risks reducing its effectiveness and the possibility of greater harm and exploitation of participants. This paper contributes towards clarifying the understanding and role of participation in AI, situating participation within its historical development, as central to contending with systems of power, as seeking forms of vibrant participation, and a as set of methods that has limitations and specific uses.

Participation in AI is a broad set of practices, but whether we use participation for algorithmic improvement, methodological innovation, or collective exploration, they can be characterised along axes of empowerment and reflexive assessment, along which we seek to move away from transactional engagements towards forms of vibrant participation that are in constant engagement with their publics and increase community knowledge and empowerment. As the case studies show, there are desirable forms of participation that are already available that we can draw inspiration from. Participation takes many different from across the AI pipeline, but for researchers, a key aim is to build the reflexive process that the probe questions hoped to initiate. New AI research and products will increasingly rely on  participation for its claims to safety, legitimacy and responsibility and we hope this paper provides part of the clarity needed to effectively incorporate participatory methods; and where we find participation lacking, we hope to provide a basis for critique, which is itself a powerful form of participation.

\begin{acks}

We would like to thank Nenad Tomašev, Sean Legassick, and Courtney Biles for the invaluable comments on an earlier version of the paper. We would also like thank the EAAMO reviewers for their feedback.  
\end{acks}

\bibliographystyle{ACM-Reference-Format}
\bibliography{EAAMO}


\appendix

\section*{Appendix}
\section{Reflexive Assessment of Participatory Practices}
\label{reflexive}

As outlined in earlier subsections, participatory efforts in AI may draw from different motivations, may have different objectives, and are characterized by different attributes that influence their effectiveness and utility for the communities they are meant to help. What makes participation meaningful may vary depending on its sociotechnical contexts, but we argue that it is critical nonetheless to maintain a deep reflexive exercise on the various objectives and characteristics of a planned participatory effort. This is particularly important when participation is called for by researchers, technologists and institutions in positions of power. 

Below, we provide a set of questions to clarify and make explicit the goals, objectives, and limitations of any particular participatory activity. 
These questions are not meant to serve as an exhaustive checklist, rather as a tool to help guide researchers and technologists towards a reflexive exercise. We include questions here that are relevant across most, if not all, contexts of AI interventions, however specific contexts might call for the inclusion of additional questions. 

\begin{enumerate}
 
\item Do the project the goals that motivate a participatory effort seek to support community interests?
\item What efforts will go into building trust with the people and communities that are involved in participatory initiatives?
\item When there is lack of trust, how is it understood in its historical and structural context (e.g., effects of racism and discrimination)? 
\item What efforts will be taken to manage and mitigate the effects of the power imbalance between the participants and the project team?
\item What efforts will be taken to manage and mitigate the effects of the power differential within the participant group(s)?
\item What efforts will be taken to ensure a transparent and open conversation between the participants and the project team?
\item What mechanism are put in place to allow participants to question the existence of the product/tool itself, rather than helping to reduce harms or improve benefits?
\item In what ways will the participation process allow for disagreement?
\item How do participants experience the process of participation?
\item What do participants own and how do they benefit?
\item In what ways is the participatory process more than a collection of individual experiences?
\item Will the participatory effort be a one-off engagement with the community, or a recurring/long-term engagement? 
\item At what step or phase of the development process will participation with communities take place? 
\item Did participants have the opportunity to refuse participation or withdraw from the process without causing direct or indirect harm to themselves or their communities?
\item Will there be flexibility in the participatory process to influence decisions made in prior phases?
\item Could the data source or curation decisions be changed, and data be re-collected based on insights from your participatory efforts?
\item Could the plans for evaluation of performance or harms be updated based on insights from your participatory efforts?
\end{enumerate}

\section{Current Modes of Participation in AI}
\label{sect:modes}

There is no single unified understanding of the term or vision of what participation is supposed to achieve. Rather, participation is a concept that encompasses a cluster of attitudes, practices, and goals, ranging from focusing on improving model performance, accuracy and data quality to building social reforms and redistribution of power and resources. A single universal definition and vision vision of what participatory AI could could become is therefore futile. In this section, we offer a characterization 
of participation in order to clearly articulate the implications and consequences of different forms of participation in ML projects. We present three broad categories of participation, not with a sharp delineation and boundaries, but as overlapping categories morphing into each other. Our characterization 
investigates participation in terms of its driving motivations and objectives.   

\subsubsection{Participation for algorithmic performance improvement}
\label{performace_improvement}

Participation of this type aims to leverage participation as a means to improve the quality, robustness, or diversity of one or more technical components of an existing system such as a dataset or model. This might include practices such as community data labeling and annotation, or collective data entry and cleaning. These kinds of practices tend to have clear and pre-defined roles, often set out by the researcher or tech developer for the ``participant'', ``modeller'' and other stakeholders. Tasks for participants tend to be discreet microtasks or one-off involvements, although they may also involve longer term engagements. 

Pre-defined and relatively rigid roles for participants enable certain kinds of objectives, but also foreclose others. These types of practices are relatively unambiguous and can have benefits such as improvement of AI model accuracy, efficiency, and robustness via improving data quality~\cite{sambasivan2021everyone}. They can also elevate the inclusion and better representation of 
subjects and communities (in datasets) that are historically marginalized in technology development. 
For instance, data annotation done by communities that are subject matter experts of certain topics can contribute to a net improvement of AI models and potentially be a net benefit to these communities themselves. But, data from marginalized communities also has the potential to be be misunderstood and misinterpreted. \citet{frey2020artificial}, for example, hired formerly gang-involved youths as domain experts that are best suited to contextualize Twitter data from same domain in order to create a more inclusive and community-informed ML systems.

Because the impact of participants is tightly scoped to particular components of a system or system behavior, overall community benefit is not guaranteed. Furthermore, the participants are often not provided with the overall picture of the larger system. Communication between developers, designers, and participants, tends to be limited and one-way, in that the participant has little room for negotiation or access to the larger goal or the downstream impact of their ``participation''. Participants may not even be aware of how their individual actions impact the overall outcomes of the system. From this perspective, participation is piecemeal and potentially extractive in the sense that it may not result in a net benefit to participants, or lead to technology that benefits those at the margins of society.

\subsubsection{Participation for process improvement} 
\label{methoological_innovation}
In other uses, participation is prioritized as a methodology that provides unique insights that can inform the overall design process (not as a means to achieve a particular pre-determined technical objective). Examples of this type include 
inclusive design, user-centered design, co-design, participatory design, human-in-the-loop model development, citizen science, and crowdsourcing. This form of participation can vary from a discreet one-off involvement to a short term recurring participation. Participation takes place in the form of activities like user feedback, surveys, focus groups, and user experience (UX) studies 
(see~\cite{lee2019webuildai, feyisa2020characterizing,jo2020lessons, ishida2007augmented}). This type of participation can be found in academic research settings, non-profit and activist organization (e.g., the participatory initiatives from the Turing Institute~\footnote{https://decidim.org/ and  https://www.turing.ac.uk/research/research-projects/citizen-science-platform-autistica} and the UN~\footnote{https://www.unodc.org/unodc/en/alternative-development/particpatoryapproaches.html} for example), as well as in industry product development settings~\footnote{https://pair-code.github.io/datacardsplaybook/}. 

This form of participation also facilitates certain types of opportunities, but also forecloses others. When done in a service or product-design setting, such participation is a means to include the beliefs, attitudes, and needs 
of customers who may have been marginalized or for whom the product is not working in their best interest. There are circumstances in which this form can benefit both the company (making profit on a better product) and the individuals and communities involved (e.g., see initiatives such as \textit{Racial Equity in Everyday Products}~\footnote{https://design.google/library/racial-equity-everyday-products/} and 
\textit{How anyone can make Maps more accessible}~\footnote{https://africa.googleblog.com/2021/05/how-anyone-can-make-maps-more-accessible.html}). While there is some flexibility to discuss, iterate and change the overall goals of a project or product, those who design and create tools often set the agenda and make major decisions. In general, this form lends itself to some level of active participation on how a project, product, or service is designed with an underlying aim of mutual benefit for the participant and those seeking participants. Still, this form of participation neither allows questions such as whether the design or service should exist at all nor goes so far as to challenge power asymmetries and oppressive structures that are at play. Participation for process improvement is more reciprocal (than participation for algorithmic improvement), but the boundaries of the project remain relatively \textit{fixed} by the designers and bigger organizational goals and participation is expected to happen within those bounds without opportunity to question or change those bounds.

\subsubsection{Participation for collective exploration}
\label{collective_exploration}

Participation of this type aims to organize a project (a practice or an activity) around the needs, goals, and futures of a particular community. Participants self-organize or are included as representatives of a community to which they belong, and the activities that constitute the means of participation are designed to generate an understanding of the views, needs, and perspectives of the group. Core concerns of communities shape the central agenda, product, or service and subsequently how participation is collectively organized or how participants are selected and engaged. This type of participation tackles question such as who participates, why, how, who benefits, what are the larger objectives of participation, as well as the downstream implications for a product or service (see~\cite{costanza2020design,katellToward2020,maori21,ahmed2020} for example). 

In this form of participation, creating, planning, decision making and building are an ongoing and emergent processes. They unfold over time, in sometimes unanticipated and unpredictable ways, through continual discussion among participants, facilitators and various stakeholders. In its extreme form, this form of participation can be mainly exploratory and a medium that enables in-depth understanding of each other and the previously unfamiliar, where people engage in active and iterative co-learning. Here, participation is not merely a means to a goal; participation is a main goal in and of itself. Participation in this form is seen as not merely a way to problem solve but rather as a rewarding and valuable activity and a process in and of itself (but also one that may result in a product or service as a by product). This form of participation is rarely found within AI. The core motivation and objectives of this form of participation are somewhat incompatible with the core values of AI research, which include efficiency, performance, accuracy, novelty, and beating state-of-the-art \cite{birhane2021values}.

Participation for collective exploration can lead to the attainment of the goals set out by other forms of participation, but also has significant implications for how a project is structured and what a project may or may not ultimately achieve. Participation as collective exploration requires prioritizing the expertise that is gained through local knowledge and lived experience. 
While this approach can accommodate both methodological innovation and fine grained objectives such as model/data improvement, such outcomes would be by-products of processes, not an originating and orienting focus. This form necessarily takes time, and may come in conflict with rapid research and design expectations. 

Decisions at every stage of the process --- from boundary definition, co-deliberation, co-conceptualization of problems, to design and development of tools/products/prototypes to data collection and annotation and everything in between --- are made through active involvement of participants. The case study `Fighting Fighting Māori data rights' provides an example of a form of activity that is participatory throughout.  
As this form of participation is driven by the needs, goals, and futures of a particular community, the need for a project, product or tool can be questioned and/or entirely discarded if it conflicts such community goals and needs. This form of participation asks not only who is/ought to participate and at what level, but also participate to what ends? Questions such as ``Is participation to enable better surveillance? To improve harmful systems? To create an anti-racist model?" form a core part of it. This form of participation, because no question is off the table, accommodates, and is sympathetic to, challenging uneven power dynamics and hierarchical and oppressive social structures.  
This form of participation is slow paced, unfolds over a long period of time, and physical space and interaction are important. It requires a significant investment in resources, money and time while the outcome may not be measurable in monetary terms (compared to participation for algorithmic performance improvement, for instance, where results are immediate). 


\end{document}